\documentstyle [12pt,a4]{article}

\newcommand{\be}{\begin{equation}}
\newcommand{\ee}{\end{equation}}
\newcommand{\bea}{\begin{eqnarray}}
\newcommand{\ena}{\end{eqnarray}}
\newcommand{\sect}[1]{\setcounter{equation}{0}\section{#1}}
\newcommand{\vs}[1]{\rule[- #1 mm]{0mm}{#1 mm}}
\newcommand{\hs}[1]{\hspace{#1 mm}}
\newcommand{\sm}[2]{\frac{\mbox{\footnotesize #1}\vs{-2}}
                   {\vs{-2}\mbox{\footnotesize #2}}}

\newcommand{\shalf}{\sm{1}{2}}
\newcommand{\al}{\alpha}
\newcommand{\var}{\varphi}

\newcommand{\cd}{\mbox{$\cal{D}$}}
\newcommand{\ck}{\mbox{$\cal{K}$}}
\newcommand{{\cg}}{\mbox{$\cal{G}$}}

\newcommand{\cc}{\mbox{$\cal{C}$}}

\newcommand{\prt}{\partial}
\newcommand{\eps}{\epsilon}

\newcommand{\und}{\underline{\mbox{N}}}

\newcommand{\r}[1]{{\cal R}_{#1}}
\newcommand{\rpi}[1]{{\cal R}_{#1}^{\pi}}
\newcommand{\wt}[1]{\widehat{#1}}
\newcommand{\sma}[1]{\mbox{\footnotesize{$ #1$}}}
\newcommand{\Z}{Z\hspace{-2mm}Z}

\newcommand{\NP}[1]{Nucl.\ Phys.\ {\bf #1}}
\newcommand{\PL}[1]{Phys.\ Lett.\ {\bf #1}}

\newcommand{\CMP}[1]{Comm.\ Math.\ Phys.\ {\bf #1}}
\newcommand{\JMP}[1]{Journ.\ Math.\ Phys.\ {\bf #1}}

\begin{document}
\renewcommand{\thefootnote}{\fnsymbol{footnote}}
\newpage
\pagestyle{empty}
\setcounter{page}{0}

\vs{20}

\begin{center}

{\LARGE {\bf Folding the $W$-algebras}}\\[2cm]

{\large L. Frappat, E. Ragoucy\footnote{Present address: NORDITA,
Blegdamsvej 17, DK-2100 Copenhagen \O, Denmark}, P. Sorba}\\
{\em{Laboratoire de Physique Th\'eorique}}
{\small E}N{\large S}{\Large L}{\large A}P{\small P}
\footnote{URA 14-36 du CNRS, associ\'ee \`a l'E.N.S. de Lyon, et au L.A.P.P.
(IN2P3-CNRS) d'Annecy-le-Vieux.}
\\
{\em Chemin de Bellevue BP 110, F - 74941 Annecy-le-Vieux Cedex,
France}\\[.5cm]

\end{center}
\vs{20}

\centerline{\bf Abstract}

\indent

In the same way the folding of the Dynkin diagram of $A_{2n}$ (resp.
$A_{2n-1}$) produces the $B_n$ (resp. $C_n$) Dynkin diagram, the symmetry
algebra $W$ of a Toda model based on $B_n$ (resp. $C_n$) can be seen as
resulting from the folding of a $W$-algebra based on $A_{2n}$  (resp.
$A_{2n-1}$). More generally, $W$ algebras related to the $B-C-D$ algebra series
can appear from $W$ algebras related to the unitary ones. Such an approach is
in particular well adapted to obtain fusion rules of $W$ algebras based on non
simply laced algebras from fusion rules corresponding to the $A_n$ case.
Analogously, super $W$ algebras associated to orthosymplectic superalgebras are
deduced from those relative to the unitary $A(m,n)$ series.

\vs{20}

\rightline{{\small E}N{\large S}{\Large L}{\large A}P{\small P}-AL-408/92}
\rightline{November 1992}

\newpage
\pagestyle{plain}

\renewcommand{\thefootnote}{\arabic{footnote}}
\setcounter{footnote}{0}
\newpage

\sect{Introduction \label{sec:1}}

\indent

The symmetries of the Dynkin diagrams of simple Lie algebras are well known to
be directly connected to the inclusion of the algebras $B_n$ in $A_{2n},C_n$ in
$A_{2n-1}, B_{n-1}$ in $D_n,G_2$ in $D_4$ and $F_4$ in $E_6$
respectively. Such a
property can sometimes be used to extend a construction based on a simply laced
algebra to ones related to non simply laced algebras. This feature has been
first, in our knowledge, pointed out and applied to the reduction of Toda field
equations in \cite{1b}. Later, it has been used to build vertex operator
representations for non simply laced algebras \cite{2b} as well as, using
symmetries of
the affine Dynkin diagrams, twisted vertex representations \cite{3b}. One may
also note that similar technics have been used to construct vertex operators of
Lie superalgebras \cite{4b}.

Coming back to \cite{1b} in which (Abelian) Toda equations relative to a
simply laced algebra are elegantly reduced to Toda equations for non simply
laced algebras, one can wonder whether such a technics can be used in the
context of $W$ algebras \cite{5b} which are symmetries of Toda models.

Actually, using the property of a Toda model based on the Lie algebra ${\cg}$
to
appear, at least at the classical level, as a constrained WZW model \cite{6b},
one knows
how to express the primary fields generating the $W$ symmetry of the model in
terms of the (constrained) current, and to deduce their Poisson brackets (P.B.)
from those
of these fundamental current fields. When restricting by folding the algebra
${\cg}$ -for example $A_{2n}$- to its folded subalgebra ${\cg}^F$ -for example
$B_n$- one will be led to make some identifications among these ${\cg}$ current
fields in order to restrict the theory to its subalgebra ${\cg}^F$. As will be
seen hereafter, a $W$ algebra relative to ${\cg}^F$ will then be explicitly
obtained and its P.B. determined from those of the previous
${{\cg}}$-$W$ algebra.
Note that such a method could be used in the case of an Abelian as well as non
Abelian Toda model based on ${{\cg}}^F$. However, although any non Abelian Toda
in
${{\cg}}^F$,subalgebra of $\cg$, can be obtained from a Toda in ${\cg}$, the
converse is not
true, and one will have to limit the folding to a subclass of non Abelian Toda
models in ${\cg}$. Let us remind that the Abelian Toda model in ${\cg}$
involves
the principal $Sl(2)$ subalgebra of ${\cg}$, while for a non Abelian model, the
corresponding $Sl(2)$ is only principal in a subalgebra of ${\cg}$.

The paper is organized as follows. We start by recalling in Section \ref{sec:2}
the folding operation on a simply laced algebra, before presenting in Section
\ref{sec:3} how
this tool can be used for connecting $W$ algebras related to Toda models. We
illustrate the method by studying in some details the different possibilities
occuring from the $A_3$-$W$ algebras (Section \ref{sec:4}).
Then the folding of
the $W$ algebra associated to the $A_n$-Abelian Toda model also denoted
$W(A_n)$ is performed, leading to the $W(B_n)$ and $W(C_n)$ algebras (Section
\ref{sec:5}). The $W(D_n)$ algebra can be obtained from a $A_{2n}$-$W$ algebra
relative to a non-Abelian Toda model (Section \ref{sec:6}) ; the case of
$W(D_4)$
which can finally be reduced to $W(G_2)$ is also considered. We note that
$B_n$-$W$ algebras can also arise from $D_{n+1}$-$W$ ones,
expliciting the Abelian
case (Section \ref{sec:7}). Then, we turn our attention to the fusion rules of
two $W$ algebras connected by the folding procedure: general relations between
their structure constants can be deduced. We can partly check our formulas on
explicit computations already performed for some specific $W$ algebras, and
also use
such results to predict new structure constant values (Section \ref{sec:8}).
In section \ref{sec:10}, we remark that a $\cg$
folding can also induce a folding on the Miura transformation
associated to $\cg$: we easily get the $B_n$
(resp. $C_n$)-Miura transformation from the $A_{2n+1}$ (resp. $A_{2n}$) one.
Of course, the method can be generalized to superalgebras, the $Sl(2)$ part
being replaced by its supersymmetric extension $OSp(1|2)$: we devote the
last section to the supersymmetric case.

\sect{Folding of a simply laced algebra \label{sec:2}}

\indent

We recall that the symmetry group $s({\cg})$ of the Dynkin diagram of a
simple Lie algebra is isomorphic to $\Z_2 = \{ -1,+1 \}$ for the $A_n, n>1$,
and
$D_n, n>5$ and $E_6$ algebras, $s_3$ for $D_4$ and finally reduces to the
identity in the other cases. We also note the isomorphisms:
\be
Out(\cg) =
\frac{Aut({{\cg}})}{Int({\cg})} \simeq \frac{Aut(\Delta_R ({\cg}))}{Int
(\Delta_R ({\cg}))} \simeq s ({\cg})
\ee
where $\Delta_R({\cg})$ denotes the root lattice of ${\cg}, Aut ({\cg})$
(resp. Int
$({\cg})$) is the group of au\-to\-mor\-phisms (resp. inner
au\-to\-mor\-phisms)
of ${\cg}$ and
$Int (\Delta_R({\cg}))$ is the Weyl group of ${\cg}$.

Denoting $\tau$ such a transformation on the simple root system $\{
\alpha_1,\alpha_2,...,\alpha_n\}$ of ${\cg}$, one gets the following table:

\vs{50}

with $\tau(\al_i) = \tau(\al_{n+1-i})$ with $i=1,2,...,n$ and $\tau^2=1$.

 \vs{50}

with $\tau(\al_i)=\al_i, \tau(\al_{n-1}) =\al_n, \tau(\al_n)=\al_{n-1}$ with
$i=1,2,...,n-2$ and $\tau^2 =1$.

\vs{50}

with $\tau(\al_2)=\al_2,\tau(\al_1)=\al_3,\tau(\al_3)=\al_4, \tau(\al_4)=
\al_1$ and $\tau^3=1$.

\vs{50}

with $\tau(\al_1)=\al_6, \tau(\al_2)=\al_5, \tau(\al_3)=\al_3,
\tau(\al_4)=\al_4$ and $\tau^2=1$.

This automorphism $\tau$ on $\Delta_R ({\cg})$ can be lifted up to a
${\cg}$-automorphism $\hat{\tau}$ by defining
\be
\hat{\tau} \left( E_{\pm \alpha_i} \right) =E_{\pm \tau (\al_i)} \ \ \ \ \
i=1,2,...,n.
\ee
with
\be
\hat{\tau} \left[ E_{\alpha_i}, E_{-\alpha_i} \right] = \tau(\alpha_i) \cdot H
\ee
the two cocycle $\eps(.,.)$ defined by
\be
\left[ E_{\alpha} , E_{\beta} \right] = \eps(\alpha, \beta) E_{\alpha+ \beta}
\ \ \ \ \  \mbox{if} \ \ \alpha + \beta  \ \ \ \ \ \mbox{is a root}
\ee
being invariant under $\tau$, i.e.:
\be
\eps \left(\tau(\alpha),\tau(\beta)\right) =\eps(\al,\beta) \ \ \ \
\alpha, \beta \in \Delta_R ({\cg})
\ee
The ${\cg}$ subalgebra invariant under the $\hat{\tau}$ automorphism can easily
be deduced: we will denote it ${\cg}^F$.
Starting from $A_{2n}$ we get the subalgebra $B_n$ generated by
\be
E_{\pm  \alpha_i} + E_{\pm \alpha_{2n-i+1}} \ \ \ \ i=1,2,...,n
\label{eq:1}
\ee
while starting from $A_{2n-1}$ one obtains the subalgebra $C_n$ generated by:
\be
E_{\pm  \alpha_i} + E_{\pm \alpha_{2n-i}} \ \ \ \ i=1,2,...,n-1 \ \ \
\ \mbox{and} \ \ \ \ E_n.
\label{eq:2}
\ee

Considering the $N \times N$ matrix representation of $Sl(N)$ the folding will
be associated to a symmetry with respect to the anti diagonal -i.e. elements
$E_{1,N}, E_{2,N-1}, ..., E_{N,1}$. Denoting as usual $E_{ij}$ the matrix with
entry 1 on the $i^{th}$ row and $j^{th}$ column, any $Sl(N)$ algebra
e\-le\-ment
will write:
\be
M=m^{ij}E_{ij} \ \ \ \ \ m^{ij} \ \ \ \ \ \mbox{being real numbers}
\ee
with the traceless condition $\sum^N_{i=1} m^{ii} =0$. In particular
the matrices
$E_{i,i+1} (i=1,2,...,N-1)$ represent the generators
$E_{\alpha_i}$
with the $\alpha_i$ constituting a set of simple roots. Then a representation
of ${\cg}^F$ will be obtained with the matrices $M$ satisfying:
\be
m^{ij}=(-1)^{i+j+1} m^{N+1-j,N+1-i}
\label{eq:2.9}
\ee

We can note that, on the upper minor diagonal, that is elements
$E_{12}, E_{23},..., E_{N-1,N}$, this condition reads:

\be
m^{i,i+1} = m^{N-i, N+1-i}
\ee
in accordance with equations (\ref{eq:1}) and (\ref{eq:2}).

On the antidiagonal, one gets:
\be
m^{i,N+1-i} = (-1)^N m^{i,N+1-i}
\ee
which implies that for $N$ even, elements on the antidiagonal are conserved,
while for $N$ odd, such elements must be put to zero.

As an example, we got from $Sl(4)$ the $Sp(4)$ algebra with general element:
\be
\left(
\begin{array}{cccc}
a & c & e & f\\
c' & b & d & -e\\
e & d' & -b & c\\
f' & -e' & c' & -a
\end{array}
\right)
\ee
and, starting from $Sl(5)$ the $SO(5)$ algebra representation:
\be
\left(
\begin{array}{ccccc}
\al & \gamma & \eps & \varphi &0\\
\gamma '&\beta&\delta&0&\varphi\\
\eps ' & \delta ' & 0 & \delta & -\eps \\
\varphi '&0&\delta ' & - \beta& \gamma \\
0 & \varphi ' & - \eps ' & \gamma ' & - \al
\end{array}
\right)
\label{eq:2.13}
\ee

Finally, from the above matrix representation of $B_n$, a natural $D_n$
representation is obtained by simply suppressing the $n^{th}$ row and the
$n^{th}$ column in the $(2n+1) \times (2n+1)$ general matrix. As a very simple
example, we can see that this restriction applied on the matrix (\ref{eq:2.13})
of $SO(5)$ yields the $SO(4)$ general matrix:
\be
\left(
\begin{array}{cccc}
\al & \gamma & \var & 0 \nonumber \\
\gamma'&\beta&0&\var \nonumber \\
\var'&0&-\beta&\gamma \nonumber \\
0&\var ' & \gamma'&-\al
\end{array}
\right)
\ee

\sect{Folding of a $W$ algebra: general conside\-ra\-tions \label{sec:3}}

\indent

A Toda model is in particular characterized by a simple Lie algebra ${\cg}$ and
an $Sl(2)$ subalgebra of ${\cg}$. This $Sl(2)$ subalgebra is itself the
principal
subalgebra of a -- up to a few exceptions -- regular subalgebra
${\ck}$ of ${\cg}$. Therefore
we will denote by $W({\cg}, {\ck})$ the $W$ symmetry algebra associated to this
model. Abelian Toda theories will correspond to the case ${\ck}={\cg}$,
with symmetry
algebra $W({\cg}, {\cg}) \equiv W({\cg})$.

\indent

Let us start from a Toda model specified by the Lie algebra ${\cg} = Sl(M)$ and
its subalgebra $\ck$ whose symmetry is therefore $W({\cg}, {\ck})$.
It is well known
that  this Toda model can be seen as a constrained WZW model with current
-- we
limit for the purpose to the right handed part, but of course, the treatment is
identical for the left handed part -- :
\be
J(x^+) = J_- + J_{>-1} (x^+)
\ee
where $J_-$ is $x$-independent and specifies the $Sl(2)$ subalgebra which is
principal in ${\ck}$. This $Sl(2)$ subalgebra defines a ${\cg}$-grading:
\be
{\cg} = \bigoplus_h {\cg}_h
\ee
and
\be
J_{\geq-1} (x^+) = \phi^{m} (x^+) \cdot X_{m} \ \ \ \ \ \ \ \ \ \ \
\ \  \mbox{with} \ \ \\
X_{m} \in \bigoplus_{h>-1} {\cg}_h
\label{eq:3.3}
\ee

By action of the residual gauge symmetry generated by ${\cg}_+$, one can
transform $J$ into:
\be
J^g = g^{-1} Jg + g^{-1} \prt_+ g
\label{eq:3.4}
\ee
with
\be
J^g = J_- + \sum_h W^{h+1} X_h
\ee
each $X_h$ being the highest weight of an irreducible representation of the
above mentioned $Sl(2)$ subalgebra in ${\cg}$. This transformation
$J \rightarrow J^g$
defines a highest weight Drinfeld-Sokolov gauge, in which the $W_{h+1} (J)$,
polynomials in the $\phi^m$ and their derivatives, show up as the primary
fields, generators of the $W({\cg},{\ck})$ algebra.

It might already be clear to the reader that we plan to deduce $W$ algebras
associated to the ${\cg}$ folded algebra ${\cg}^F$ by identifying some of the
fields $\phi^{m}$ in (\ref{eq:3.3}) in order for the current $J$
to be expressed in
terms of ${\cg}^F$ elements.

Of course, any $W({\cg},{\ck})$ cannot give rise to a $W({\cg}^F,{\ck}')$.
Indeed, any
$Sl(2)$ in ${\cg}^F$ is an $Sl(2)$ in ${\cg}$ since ${\cg}^F$ is a
${\cg}$-subalgebra,
but the converse is not true. Therefore, starting from ${\cg}=A_{N-1}$, one has
first to determine in ${\cg}$ the $Sl(2)$ subalgebras which will "survive" the
folding. A simple way to select these $Sl(2)$ subalgebras of ${\cg}^F$ in
${\cg}$
is to compute the decomposition with respect to each $Sl(2)$ in ${\cg}$ of the
$\underline{\mbox{N}}$-dimensional fundamental ${\cg}$-representation.
Then to do the
same on $\underline{\mbox{N}}$ for each $Sl(2)$ in ${\cg}^F$, and
to compare the two
obtained lists (see for example \cite{7} for many examples). Each time the same
decomposition will appear in the two lists, one could conclude that the
corresponding $Sl(2)$ in ${\cg}$ can be seen as an $Sl(2)$ in ${\cg}^F$, and
vice
versa. The proof is simply based on the property of the $\und$ of $A_{N-1}$
to reduce with respect to its singular subalgebra $B_{\frac{N-1}{2}}$ if $N$ is
odd, or $C_{N/2}$ if $N$ is even, to $\und$ itself. Let us remark {\em en
passant}
that for
$N$ odd, since the representation $\und$ of $B_{\frac{N-1}{2}}$ is real, the
allowed $Sl(2)$ decompositions:
\be
\und = \bigoplus_j n_j \cd_j
\ee
are  such that $n_j$ must be even when $j \in \Z + \shalf$. It is the
opposite for $C_{\frac{N}{2}} $ ($N$ being then even): $n_j$ must be even
for $j \in \Z$.

We illustrate our method with $A_3$, which folds into $C_2$. In this case, we
can easily regognize the ${\cg}$ regular subalgebra $\ck$ in which $Sl(2)$ is
principal and its folded image ${\ck}^F$ in ${\cg} ^F$ (the situation looks
more
complicated when the rank is higher).
\be
\begin{array}{cccc}
Sl(2) \ \mbox{decompositions} & A_3 \ \mbox{Regular subalgebra} && B_2
\ \mbox{regular subalgebra} \nonumber \\
& \mbox{where} \ \ Sl(2) \ \ \mbox{is principal} && \ \nonumber \\
\underline{4} =\cd_{3/2} & \begin{picture}(70,15)
\thinlines
\put(5,5){\circle{10}}
\put(11,5){\line( 1, 0){ 15}}
\put(32,5){\circle{10}}
\put(38,5){\line( 1, 0){ 15}}
\put(59,5){\circle{10}}
\end{picture}
=A_3 & \rightarrow & \begin{picture}(42,15)
\thinlines
\put(5,5){\circle{10}}
\put(11,4){\line( 1, 0){ 15}}
\put(11,6){\line( 1, 0){ 15}}
\put(32,5){\circle*{10}}
\end{picture}
=B_2 \\
2\cd_{1/2} & \bigcirc \ \bigcirc =2A_1 & \rightarrow & \bigcirc =A^2_1 \\
\cd_{1/2} + 2\cd_0 & \bigcirc =A_1 & \rightarrow &
\begin{picture}(10,15) \put(5,5){\circle*{10}}\end{picture} = A_1 \\
\cd_1 + \cd_0 & \begin{picture}(42,15)
\thinlines
\put(5,5){\circle{10}}
\put(11,5){\line( 1, 0){ 15}}
\put(32,5){\circle{10}}
\end{picture}
=A_2 & \rightarrow & \ \mbox{no counterpart}.
\end{array}
\ee

Now that we have selected the $W(A_{N-1}, {\ck})$ which can undergo the
foldering
procedure, it is sufficient for obtaining the corresponding $W$ algebra to
constrain, inside the primary field realization $W_{h+1} (\phi^m)$, the
components $\phi^m$ of (\ref{eq:3.3}) reexpressed in
the canonical ${\cg}$-basis:
\be
\phi^m X_m = \var^{ij} E_{ij}
\ee
as follows:
\be
\var^{ij} = (-1)^{i+j+1} \var^{N+1-j, N+1-i}
\label{eq:3.9}
\ee
that is in accordance with formula (\ref{eq:2.9}).

It is rather clear that we would have obtained the same $W$ fields starting
from ${\cg}^F$ and its $Sl(2)$ subalgebra, itself principal in ${\ck} \subset
{\cg}$, and using
the gauge fixing expressed above (cf. eq.(\ref{eq:3.4})). Some of
the modified $W$'s might
become identically zero if they do not correspond to an highest weight in the
reduction of ${\cg}^F$ with respect to the $Sl(2)$ under consideration.
Some others might
also be linearly dependent: we make explicit such features by determining in
the next section the $W$ algebras arising from the $A_3$ folding.

Finally, let us remark that the P.B. of such modified $W_{h+1}$'s
can naturally be
deduced from the P.B. of the original $W_{h+1}$'s, themselves being
calculated from
the P.B. of the constituting current fields $\phi^m$. In particular the
vanishing
of a $W_{h+1}$ field by imposing condition (\ref{eq:3.9}) implies this
field to disappear in
the right hand side of the P.B. of the folded $W$ algebra.

\sect{Example: $C_2$-$W$ algebras from $A_3$-$W$-ones \label{sec:4}}

\indent

Let us treat separately  each of the three admissible $Sl(2)$ embeddings
determined in the previous section:

\indent

i) $\underline{4}={\cd}_{3/2} \ \ \ \mbox{or Abelian case}$ the
$Sl(2)_{ppal}$ is generated by:
\bea
M_- &=& E_{21} + E_{32} + E_{43}, \nonumber\\
M_+ &=&
3E_{12} + 4E_{23} + 3 E_{34}, \nonumber\\
M_0 &=& \sm{3}{2} E_{11} + \shalf E_{22} - \shalf
E_{33} - \sm{3}{2} E_{44}. \nonumber
\ena

The corresponding grading can be read on the following antisymmetric
matrix (i.e. $g_{ij}$ being the grade of the generator $E_{ij}$).
\be
g=
\left(
\begin{array}{cccc}
0&1&2&3\\
\ &0&1&2\\
(-)&&0&1\\
&&&0
\end{array}
\right)
\ee
Determining the $A_3$-highest weights under $Sl(2)_{ppal}$ implies:
\be
J^g =
\left(
\begin{array}{cccc}
0&3W_2&W_3&W_4\\
1&0&4W_2&W_3\\
0&1&0&3W_2\\
0&0&1&0
\end{array}
\right)
\ee
Then, imposing the constraint (\ref{eq:3.9}) and therefore $J^g$ to
belong to $B_2$ implies immediately:
\be
W'_3 = - W'_3
\ee
where the $'$ indicates that condition (\ref{eq:3.9}) is applied.

As could be expected the $W$ algebra symmetry of the Abelian Toda model
for $A_3$,
and generated by $W_2, W_3, W_4$ of conformal spin 2,3,4, reduces to the
the $B_2$-$W$ algebra generated by $W'_2$ and $W'_4$.

\indent

ii) $\underline{4} =2{\cd}_{1/2}$ Here, we get for
\bea
M_-&=&E_{32} + E_{41}, M_+ = E_{14} + E_{23}\nonumber \\
\mbox{and } M_0 &=& \shalf E_{11} + \shalf E_{22} - \shalf E_{33} - \shalf
E_{44}.\nonumber
\ena
The grading reads:
\be
g=
\left(
\begin{array}{cccc}
0&0&1&1\\
&0&1&1\\
(-)&&0&0\\
&&&0
\end{array}
\right)
\ee
One gets three spin 1 primary fields $W^{(i)}_1 \ i=1,2,3$, and four spin 2
fields $W^{(j)}_2$ $j=1,2,3,4.$
\be
J^g=
\left(
\begin{array}{cccc}
W^{(1)}_1 & W^{(2)}_1 & W^{(1)}_2 & W^{(2)}_2 \\
W^{(3)}_1 & -W^{(1)}_1 & W^{(3)}_2 & W^{(4)}_2 \\
0 & 1 & -W^{(1)}_1 & W^{(3)}_1 \\
1 & 0 & -W^{(2)}_1 & W^{(1)}_1
\end{array}
\right)
\longrightarrow
\left(
\begin{array}{cccc}
0 & W'_1 & W'_2 & W''_2 \\
W'_1 & 0 & W'''_2 & -W'_2 \\
0 & 1 & 0 & W'_1\\
1 & 0 & W'_1 & 0
\end{array}
\right)
\ee
the folding implies:
\be
W^{(1)'}_1= 0 \ \ ; \ \ W_1^{(2)'} = W_1^{(3)'}  ; W^{(1)'}_2 = - W^{(4)'} _2
\ee
which means in $B_2$ only one spin 1 field $W'_1$ and three spin two fields
$W'_2,W''_2,W'''_2$.

\indent

iii) $\underline{4}={\cd}_{1/2} + 2{\cd}_0$. Then, one has:
$M_- = E_{41}, \ M_+ = E_{14}, \ M_0 = \shalf E_{11} - \shalf
E_{44}$. The grading reads:
\be
g=
\left(
\begin{array}{cccc}
0 & \shalf & \shalf & 1 \\
&0&0&\shalf \\
(-)&&0&\shalf \\
& & & 0
\end{array}
\right)
\ee
and provides one spin 2, four spin $3/2$ and four spin 1 $W$-fields:
after folding, one gets three spin 1 and two spin $3/2 \ W$ fields.
\be
\left(
\begin{array}{cccc}
0 & W^{(1)}_{3/2} & W^{(2)}_{3/2} & W_2 \\
0 & W^{(1)}_1 & W^{(2)}_1 & W^{(3)}_{3/2} \\
0 & W^{(3)}_1 & -W^{(1)}_1 & W^{(4)}_{3/2} \\
1 & 0 & 0 & 0
\end{array}
\right)
\longrightarrow
\left(
\begin{array}{cccc}
0 & W'_{3/2} & W''_{3/2} & W'_2 \\
0 & W'_1 & W''_1 & -W''_{3/2} \\
0 & W'''_1 & -W'_1 & W'_{3/2} \\
1 & 0 & 0 & 0
\end{array}
\right)
\ee

\sect{$B_n$- and $C_n$-$W$ algebras from $A_N$-$W$ algebra (the Abelian case)
\label{sec:5}}

\indent

It is well known that in the Abelian Toda model based on the Lie algebra
${\cg}$,
the primary fields $W_s$ of the symmetry $W$ algebra are of conformal spin
$s=2,3,...n+1$ when ${\cg}=A_n$, and $s=2,4,..,2n$ when ${\cg}=B_n$ or $C_n$.

One can easily check by direct computation that starting form $A_{N-1}$ and
applying our folding technics will imply for the odd conformal spin $W$
generators to vanish.

First, the $Sl(2)_{ppal}$ subalgebra in $A_{N-1}$ can be seen as generated by:
\be
M_- = \sum^{N-1}_{i=1} E_{i+1,i} \ \ \ M_0=\sum^N_{i=0} \shalf (N+1 - 2i)
E_{ii} \ \ \ M_+ = \sum^{N-1}_{i=1} \shalf i(N-i) E_{i,i+1}
\ee

The $A_{N-1}$ Lie algebra decomposes under the adjoint action of $Sl(2)_{ppal}$
into the sum of representations:
\be
{\cg} = \bigoplus_{p=1}^{N-1} {\cd}_p
\ee

In the $Sl(2)$ representation ${\cd}_p$, the highest weight $M_p$ reads:
\be
M_p = \sum^{N-p}_{j=1} a^j_p E_{j,j+p} \ \ \ \mbox{with}\ \ \
a^j_p = \frac{(j+p-1)!(N-j)!}{(j-1)! (N-p-j)!}
\ee

\indent

One notes the obvious symmetry:
\be
a^j_p = a_p^{N+1-j-p}
\ee

Now, let us apply the folding condition (\ref{eq:2.9}), that is
\be
a^j_p = (-1)^{p+1} \ a^{N+1-j-p}_p
\ee

The compatibility of these two last relations implies the vanishing of the
$M_p$ generators when $p$ is even, that is  since $M_p$ will be associated with
the field $W_{p+1}$, the vanishing of the odd conformal spin fields $W_{2k+1}$.

\sect{$D_n$-$W$ algebras from $A_{2n}$-$W$ algebras \label{sec:6}}

\indent

Although $D_n$ is simply laced, it is a maximal (regular) subalgebra of $B_n$.
Therefore, one can think of deducing the $W$-algebras associated to $D_n$-Toda
models starting from $A_{2n}$-Toda symmetry.

Limiting  ourselves to the construction of the $W$-symmetry of the Abelian Toda
model relative to $D_n$, one will have to consider the non Abelian Toda model
based on $A_{2n}$ such that the $Sl(2)$ which determines the grading is
principal in $A_{2n-2}$. Indeed, the $D_n$ fundamental representation of
dimension $2n$ reduces with respect to the $Sl(2)$ principal in $D_n$ as:
\be
\underline{2n} = {\cd}_{n-1} \oplus {\cd}_0
\ee
Therefore the only possibility is to find out an $Sl(2)$ in $A_{2n}$ such that
the $\underline{2n+1}$ representation reduces with respect to it as:
\be
\underline{2n+1} = {\cd}_{n-1} \oplus 2{\cd}_0.
\ee
This $Sl(2)$ is principal in $A_{2n-2}$.

The $A_{2n}$ adjoint representation decomposes with respect to the
$Sl(2)_{ppal}$ in $A_{2n-2}$ as :
\bea
&& \left( {\cd}_{n-1} \oplus 2{\cd}_0 \right) \times \left( {\cd}_{n-1} \oplus
2{\cd}_0 \right) - {\cd}_0  \nonumber \\
&& = {\cd}_{2n-2} \oplus {\cd}_{2n-1} \oplus ... \oplus {\cd}_{n-2} \oplus
5{\cd}_{n-1} \oplus {\cd}_{n} \oplus ... \oplus {\cd}_{1} \oplus 4{\cd}_{0}
\ena
leading to a set of $W_s$ primary fields constituted by one field $W_s$ of
conformal spin $s=2n-1, 2n,...n-1,n+1,...1$ respectively, five with $s=n$
and four with $s=1$.

One will take in the $(2n+1) \times (2n+1)$ matrix representation the $A_1$
commuting with $A_{2n-2}$ generated by $E_{n-1}-E_{n+1}, E_{n-1,n+1}$ and
$E_{n+1, n-1}$. The grading $H$ will be represented by
\be
H=diag \left[ 2n-1, 2n-2, ..., 1,0,0,0,-1,-2,...,-(2n-1) \right]
\ee
and the grading by
\be
\left(
\begin{array}{cccccccccccccc}
0 &1 &2 &\cdots &n-2 &n-1 &n-1 &n-1 &n &\cdots &\cdots &2n-3 &2n-2
\nonumber \\
 &\ddots &\ddots &\ddots & &n-2 &n-2 &n-2 &\ddots &\ddots & & &2n-3 \nonumber
\\
 & &\ddots &\ddots &\ddots &\vdots &\vdots &\vdots & &\ddots &\ddots & &\vdots
\nonumber \\
 & & &0 &1 &2 &2 &2 & & &\ddots &\ddots &\vdots \nonumber \\
 & & & &0 &1 &1 &1 &2 & & &\ddots &n \nonumber \\
 & & & & &0 &0 &0 &1 &2 &\cdots &n-2 &n-1 \nonumber \\
 & & & & &0 &0 &0 &1 &2 &\cdots &n-2 &n-1 \nonumber \\
 & & & & &0 &0 &0 &1 &2 &\cdots &n-2 &n-1 \nonumber \\
 & & & & & & & &0 &1 &2 &\cdots &n-2 \nonumber \\
 & & & & &(-) & & & &\ddots &\ddots & &\vdots \nonumber \\
 & & & & & & & & & &\ddots &\ddots &2 \nonumber \\
 & & & & & & & & & & &\ddots &1 \nonumber \\
 & & & & & & & & & & & &0 \nonumber \\
\end{array}
\right)
\ee

The $Sl(2)_{ppal}$ in $A_{2n-2}$ reads in our basis:
\bea
M_0 &=& H\ =\ \sum^n_{i=1} (n-i) (E_{ii} - E_{2n+2-i,2n+2-i} ) \nonumber \\
M_- &=& \sum^{n-1}_{i=1} (E_{i+1,i} + E_{2n+2,i,2n+1-i}) +
E_{n+2,n-1} + E_{n+3,n}
\nonumber \\
M_+ &=& \sum^{n-2}_{i=1} \frac{i(2n-1-i)}{2}  (E_{i,i+1} + E_{2n+1-i,2n+2-i})
\nonumber\\
&& + \frac{n(n-1)}{4} (E_{n-1,n} + E_{n-1,n+2} + E_{n,n+3} + E_{n+2,n+3})
\ena

The general expression for the highest weight $M_p$ in ${\cd}_p$ takes a rather
cumbersome form:
\bea
M_0^{(1)} &=& \sum^{n-1}_{j=1} \left( E_{jj} + E_{2n+2-j,2n+2-j}\right) -2(n-1)
E_{n+1,n+1} + (E_{n,n+2} + E_{n+2,n}) \nonumber \\
M^{(2)}_0 &=& (E_{n,n} + E_{n+2,n+2}) - 2E_{n+1,n+1}
-(E_{n,n+2}+E_{n+2,n}) \nonumber \\
M_0^{(3)} &=& E_{n,n+1} - E_{n+2,n+1} \nonumber \\
M_0^{(4)} &=& E_{n+1,n+2} - E_{n+1,n}  \nonumber \\
M_p &=& \sum^{n-p-1}_{j=1} 2d^j_p \left( E_{j,j+p} +
E_{2n+2-j-p,2n+2-j} \right) + \sum^n_{j=n-p} d^j_p E_{j,j+p+2} \nonumber \\
&& + d^{n-p}_p (E_{n-p,n} + E_{n+2,n+p+2} )\ \ \ 1 \leq p<n-1  \nonumber \\
M^{(1)}_{n-1} &=& 2d^1_{n-1} E_{1,n+2} + \sum^{n-1}_{j=2} d^j_{n-1} E_{j,
n+1+j} + 2d^n_{n-1} E_{n,2n+1}  \nonumber \\
M^{(2)}_{n-1} &=& E_{1,n} - E_{1,n+2} \nonumber \\
M^{(3)}_{n-1} &=& E_{1,n+1} \nonumber \\
M^{(4)}_{n-1} &=& E_{n+1,2n+1} \nonumber \\
M^{(5)}_{n-1} &=& E_{n+2,2n+1} - E_{n,2n+1} \nonumber \\
M_{n-1+p} &=& \sum^{n-p}_{j=1} d^j_{n-1+p} E_{j,j+n+p+1}
\ena
with
\be
d^j_q = \frac{(j+q-1)!(2n-1-j)!}{(2n-1-q-j)!(j-1)!}
\ee

Note in particular:
\be
d^j_q = d_q^{2n-j-q}
\label{eq:3}
\ee

Performing once more the folding $A_{2n} \rightarrow B_n$ according to
(\ref{eq:2.9}) (and using intensively (\ref{eq:3})) we get the highest
weights in $B_n$
\bea
&&M_0({B_n}) = M^{(3)}_0(A_{2n}) + M^{(4)}_0(A_{2n})\nonumber \\
&&M_{2r+1}({B_n}) = M_{2r+1}({A_{2n}}) \ \ \ \ r \in \{0,1,...,n-2\}
\ \ \mbox{but}\ \ 2r+1\neq n-1\nonumber
\\
&&M_{n-1}^{(1)}({B_n}) = M^{(2)}_{n-1}(A_{2n}) + (-1)^n M^{(5)}_{n-1}(A_{2n})
\nonumber \\
&&M^{(2)}_{n-1}(B_n) = M^{(3)}_{n-1}(A_{2n}) - (-1)^n M^{(4)}_{n-1}(A_{2n})
\nonumber \\
&&M^{(3)}_{n-1}(B_n) = \left\{ \begin{array}{l}
\sum^{n-1}_{j=2} d^j_{n-1} E_{j,n+1+j} + 2 d^1_{n-1} \left( E_{1,n+2} +
E_{n,2n+1} \right) \ \ \mbox{when } n \ \mbox{is even} \nonumber \\
0 \ \ \mbox{when } n \ \mbox{is odd}.
\end{array}
\right.
\ena

In summary, we get for this $B_n$ non Abelian Toda symmetry, one $W$ generator
of conformal spin $s=1$, one of spin $s=2,4,...,2n-2$ respectively and in
addition two others of spin $s=n$. This result is in accordance with the
adjoint
decomposition of $B_n$ with respect to the $Sl(2)$ principal in $B_{n-1}$
obtained by folding of $A_{2n-2}$:
\be
({\cd}_{n-1} \oplus 2{\cd}_0 )^2_A = \oplus^{n-1}_{j=1} {\cd}_{2j-1} \oplus 2
{\cd}_{n-1} \oplus {\cd_0}
\ee

Finally, we reach the $D_n-W$ symmetry by suppressing in the
$(2n+1)\times(2n+1)$ matrix representation the $(n+1)^{th}$ row and
$(n+1)^{th}$ column (cf. end of section \ref{sec:2}).
Then, we immediately see that
$M_0({B_n})$ disappears as well as $M^{(2)}_{n-1}(B_n)$. We are led, as
expected,
to the $D_n$ Abelian Toda $W$-symmetry generated by $W_2, W_4,...W_{2n-2},W_n$.
We remind the adjoint decomposition of $D_n$ with respect to the $Sl(2)$
principal in $D_n$:
\be
({\cd}_{n-1} \oplus {\cd}_0 )^2_A = \oplus^{n-1}_{j=1} {\cd}_{2j-1} \oplus
{\cd}_{n-1}
\ee

Let us illustrate this construction on the case of $A_8$. After getting $D_4$,
another folding will produce the $G_2$ exceptional algebra.

Using the above general formulas, the grading we choose is:
\be
g=
\left(
\begin{array}{ccccccccc}
0&1&2&3&3&3&4&5&6\nonumber \\
&0&1&2&2&2&3&4&5 \nonumber \\
&&0&1&1&1&2&3&4 \nonumber \\
&&&0&0&0&1&2&3 \nonumber \\
&&&0&0&0&1&2&3 \nonumber \\
(-)&&&0&0&0&1&2&3 \nonumber \\
&&&&&&0&1&2\nonumber \\
&&&&&&&0&1\nonumber \\
&&&&&&&&0
\end{array}
\right)
\ee
and the $Sl(2)$ principal in $A_6$ we use is generated by:
\bea
M_0 &=& H=3(E_{11}-E_{99})+2(E_{22} -E_{88}) + (E_{33} -E_{77}) \nonumber \\
M_- &=& (E_{21} + E_{98}) +(E_{32} + E_{87}) + (E_{43} +E_{76}) + (E_{63} +
E_{74}) \nonumber \\
M_+ &=& 3(E_{12} + E_{89}) + 5(E_{23} + E_{78}) + 3(E_{34} + E_{67}) + 3(E_{36}
+ E_{47})
\ena
{}From the decomposition of the adjoint representation of $D_4$ with respect to
$Sl(2)$, we deduce the highest weights:
\bea
M_1({D_4}) &=& M_+ \nonumber \\
M^{(1)}_3(D_4) &=& E_{14}-E_{16} -E_{49} + E_{69} \nonumber \\
M^{(2)}_3(D_4) &=& E_{14} + E_{16} + 2(E_{27} + E_{38}) + E_{49} + E_{69}
\nonumber \\
M_5({D_4}) &=& E_{18} + E_{29}
\ena
leading to a $W$ algebra with primary fields of spin $s=2,4,4,6$ as expected in
the $D_4$ Abelian Toda model.

Now, we want to fold $D_4$ into $G_2$

\vs{50}

In our notations, one can use for the generator associated to
\bea
\al_1 &:& E_{12} + E_{89} \nonumber \\
\al_2 &:& E_{23} + E_{78} \nonumber \\
\al_3 &:& E_{34} + E_{67} \nonumber \\
\al_4 &:& E_{36} + E_{47}
\ena
that is, after folding, the generator associated to the short root $\al$ of
$G_2$ will be:
\[
(E_{12}+E_{89}) + (E_{34}+E_{67}) + (E_{36} + E_{47})
\]

That will imply, on the $9\times 9$ matrix representation $M=[m^{ij}]$
\be
m^{12} = m^{34} = m^{36} = m^{89} = m^{67} = m^{47}
\ee
the other constraints on $m^{ij}$ will be fixed by performing the
commutation relations from the
generators associated to the simple roots; for example, we will get:
\bea
-m^{13} &=& m^{24} = m^{26} \ ; \ m^{14} = m^{16} = -\shalf m^{38}
\nonumber \\
m^{21} &=& m^{43} = m^{63} \ ; \ -m^{31} = m^{42} = m^{62} \ ; \ m^{41} =
m^{61}
= - \shalf m^{83}
\ena

Then, will only survive as highest weights in the $G_2$ adjoint representation
decomposed with respect to its principal $Sl(2)$
\be
M_1({D_4}) = M_1({G_2}) = M_+ \ \ \ \mbox{and} \ \ \ M_5({D_4}) = M_5({G_2})
\ee
as can be expected. In other words, we have got the $W_2$ and $W_6$ primary
fields generating the $W$ symmetry of the $G_2$-Abelian Toda model from the
$W(A_8, A_6)$ algebra, using the notations introduced at the begining of
section \ref{sec:3}.

\sect{$B_{n-1}$-$W$ algebras from $D_n$-$W$ algebras \label{sec:7}}

\indent

Of course, one may also think about getting $W$ algebras associated to
$B_{n-1}$
from the folding $D_n \rightarrow B_{n-1}$. Let us hereafter illustrate the
Abelian case.

We use the $D_n$ matrix representation already studied in Section \ref{sec:6},
that is we start with $(2n+1)\times(2n+1)$ matrices with entries satisfying the
antidiagonal symmetries specified by (\ref{eq:2.9}), and we ignore the
$(n+1)^{th}$
row as well as the $(n+1)^{th}$ column. We can associate to a simple root
system
$\{ \al_1,...,\al_n\}$ the following generators:
\be
E_{\al_i} \equiv E_{i,i+1} + E_{2n+1-i,2n+2-i} \ \ \ \ \ i=1,2,...,n-1
\ee
and
\be
E_{\al_n} \equiv E_{n-1,n+2} + E_{n,n+3}.
\ee

The folding $D_n \rightarrow B_{n-1}$ will imply to replace the two last
generators into their sum:
\be
E_{\al_{n-1}} + E_n \equiv E_{n-1,n} + E_{n+2,n+3} + E_{n-1,n+2} + E_{n,n+3}
\ee
then, the remaining $B_{n-1}$ generators $m^{ij} E_{ij}$ will satisfy, in
addition with (\ref{eq:2.9}), the condition:
\be
\begin{array}{ll}
m^{j,n} = m^{j,n+2} & \nonumber \\
 & j \in \{1,...,n\} \cup \{n+2,...,2n+1\} \label{eq:7.2} \\
m^{n,j} = m^{n+2,j} & \nonumber
\end{array}
\ee

Now let consider the consequences of this last condition on the highest weight
of $D_n$. We rewrite their explicit expressions already obtained in section
\ref{sec:6} for the Abelian case: in other words, the highest weights of the
$Sl(2)_{ppal}$ representations in the $D_n$ adjoint representation can be
chosen as:
\bea
M_{2q-1}({D_n}) &=& \sum^{m-2q}_{j=1} 2d^j_{2q-1} \left( E_{j,j+2q-1} +
E_{2n+3-j-2q,2n+2-j} \right) + \sum^n_{j=n+1-2q} d^j_{2q-1} E_{j,j+2q+1}
\nonumber \\
&& + d^{n-2q+1}_{2q-1} \left( E_{n+1-2q,n} + E_{n+2,n+2q+1} \right) \nonumber
\\
&& \ \nonumber \\
&& \ \ \ \ \ \ \ \ \ \ \ \ \ \ \ \mbox{with } q \ \mbox{such that: } 0<2q-1<n-1
\nonumber \\
&& \ \nonumber \\
M_{n-1}'(D_n) &=& \left( E_{1,n} - E_{1,n+2} \right) + (-1)^n \left(
E_{n+2,2n+1} - E_{n,2n+1} \right) \\
&& \ \nonumber \\
M_{n-1}''({D_n}) &=&
\left\{\begin{array}{l}
\sum^{n-1}_{j=2} d^j_{n-1} E_{j,n+1+j} + 2d^1_{n-1} \left(
E_{1,n+2} + E_{n.2n+1} \right)
\ \mbox{when } n \ \mbox{is even} \nonumber \\
0 \ \ \ \ \ \mbox{when } n \ \mbox{is odd} \nonumber
\end{array}\right.\\
&& \ \nonumber \\
M_{2q-1}({D_n}) &=& \sum^{2n-2q}_{j=1} d^j_{2q-1} E_{j,j+2q+1} \ \ \ \ \
\mbox{with } q \ \mbox{such that} \ n-1<2q-1<2n-2 \nonumber
\ena

Condition (\ref{eq:7.2}) joined to the property (\ref{eq:3}) on the $d^j_p$
coefficients leads directly to the results:
\bea
M_{2q-1}({B_{n-1}}) \equiv M_{2q-1}({D_n}) &\mbox{with}& 0<2q-1<2n-2
\nonumber\\
&\mbox{and}& 2q-1 \neq n-1
\ena
and, when $n$ is even only:
\be
M_{n-1}({B_{n-1}}) \equiv M_{n-1}'({D_n}) + M_{n-1}''({D_n}) =
\sum^n_{j=1}d^j_{n-1} E_{j,n+1+j} + d^1_{n-1} \left( E_{1,n} + E_{n+2,2n+1}
\right)
\ee
in accordance with the primary field spin content $s=2,4,...,2n-2$ of the
$W(B_{n-1},B_{n-1})$ algebra.

\sect{Properties of the fusion rules \label{sec:8}}

\indent

As already discussed in Section \ref{sec:2} the P.B. of the $W({\cg}^F,
{\ck}^F)$
algebra can directly be deduced from the P.B. of the $W({\cg},{\ck})$ algebra.
Indeed, one has just to make in the $W({\cg},{\ck})$ primary fields, the
identifications among the current fields as expressed for example in
(\ref{eq:2.9})
for the folding $A_{2n} \rightarrow B_n$ or $A_{2n-1} \rightarrow C_n$, in
order to obtain the $W({\cg}^F,{\ck}^F)$ primary fields, and to keep
the same P.B.
In these identifications, some of the $W$-fields may vanish. That is the case
of any primary field with an odd conformal spin in the folding
\[
W(A_{2n}) \rightarrow W(B_n) \ \ \ \ \mbox{ or } \ \ W(A_{2n-1})
\rightarrow W(C_n)
\]
(we recall the notation $W({\cg},{\cg}) \equiv W({\cg})$)

In the folding $W(D_n)\rightarrow W(B_{n-1})$ the $W_n$ field vanishes when
$n$ is odd, while for $n$ even, the two fields $W_n$ and $W'_n$ become
identical. In all these cases of "Abelian" folding each primary field
in $W({\cg})$
with even conformal spin becomes a primary field of $W({\cg}^F)$.

Denoting $C^k_{ij}({\cg})$ the general structure constant of the fusion rule
\be
{[W_i]} \cdot {[W_j]} = \delta_{i,j}\,\frac{c}{i}\,{[I]}\,
+\, {\cc}^k_{ij}({\cg}) {[W_k]}
\label{eq:8.1}
\ee
for the $W({\cg})$ algebra, the folding properties imply:
\bea
C^k_{ij} (B_n) &=& C^k_{ij} (A_{2n}) \nonumber \\
C^k_{ij} (C_n) &=& C^k_{ij} (A_{2n-1}) \nonumber \\
C^k_{ij} (D_n) &=& C^k_{ij} (A_{2n-2}) \ \ \ \ \ i,j,k \neq n.
\label{eq:8.2}
\ena
Note that this last relation results from the two foldings $D_n \rightarrow
B_{n-1}$ and $A_{2n-2} \rightarrow B_{n-1}$.

It is interesting to remark that the square of the structure constant
$C^4_{44}$ has been explicitly computed \cite{H1} for the quantum algebra
$W(A_n)$, i.e.:
\be
{[W_4]} \cdot {[W_4]} = \frac{c}{4} {[I]} + C^4_{44}(A_n) {[W_4]} +...
\label{eq:8.3}
\ee
$C^4_{44} (A_n)$ is a function of $n$ as well as of the central extension $c$.
In order to obtain its value for the classical $W(A_n)$ algebra, one has to
take the $c \rightarrow \infty$ limit, that is:
\be
\left[ C^4_{44} (A_{n-1}) \right]^2 = \frac{36}{5}
\frac{(n^2-19)^2}{(n^2-4)(n^2-9)}
\label{eq:8.4}
\ee

Then, owing to relation (\ref{eq:8.2}), we have:
\be
\left[ C^4_{44} (B_n) \right]^2 = \left[ C^4_{44} (A_{2n}) \right]^2 =
\frac{36}{5} \frac{(2n^4 + 2n -9)^2}{(2n+3)(2n-1)(n-1)(n+2)}
\label{eq:8.5}
\ee
and
\be
\left[ C^4_{44} (C_n) \right]^2 = \left[ C^4_{44} (A_{2n-1}) \right]^2 =
\frac{9}{5} \frac{(4n^2 -19)^2}{(n^2-1)(4n^2-9)}
\label{eq:8.6}
\ee

In particular, we get, always in the classical limit:
\be
\left[ C^4_{44} (B_2) \right]^2 = \left[ C^4_{44} (C_{2}) \right]^2 =
\left[ C^4_{44} (A_4) \right]^2 = \left[ C^4_{44} (A_3) \right]^2 =
\frac{27}{35}
\label{eq:8.7}
\ee
in accordance with the isomorphism $B_2 \cong C_2$, our folding property, and
the numerical value obtained for $W(B_2) = W(C_2)$ in ref. \cite{KW}. Note that
in this just quoted paper, the authors also calculated the square of the
$C^4_{44}$ coefficient for two of the $W$ algebras with primary fields 2,4,6,
that
is for $W(B_3)$ and $W(C_3)$. By rel. (\ref{eq:8.2}) and (\ref{eq:8.5}) we get:
\be
\left[ C^4_{44} (B_3) \right]^2 = \frac{18}{5} \ \ \ \ \left[ C^4_{44} (C_3)
\right]^2 = \frac{289}{120}
\label{eq:8.8}
\ee

These values differ by a vector $\shalf$ from the values given in
\cite{KW}. A direct check \cite{H2} of the computations performed in
\cite{KW} let appear that indeed a factor $\shalf$ is missing in the last
formula of section \ref{sec:6}: the given expression of
${\kappa}^2 = f(c) \pm g(c)$ has to be replaced by:
${\kappa}^2 = \shalf [f(c) \pm g(c)]$.
Note that we
can also add the following information: the sign "+" in ${\kappa}^2$
corresponds to
$W(B_3)$, while the sign "$-$" to $W(C_3)$.

Finally, the study in \cite{KW} of the fusion rules for the (unique)
algebra admitting in addition to the stress energy tensor, one  primary field
of spin $s=6$, that is the algebra $W(G_2)$, allows to deduce:
\be
\left[ C^6_{66} (A_6) \right]^2 = \left[ C^6_{66} (D_4) \right]^2
= \left[ C^6_{66} (B_3) \right]^2 = \left[ C^6_{66} (G_2) \right]^2
= \frac{1352}{2079}.
\ee

\sect{Remark on the Miura transformation \label{sec:10}}

\indent

As an exercise, let us check our folding approach in the case of the free
massless Bose field $\phi$ representation of some $W$ algebra. The $W_2$ and
$W_3$ primary fields governed by the $Sl(3)$ symmetry can then be expressed as:
\bea
W_2 &=& - \frac{1}{4} : (\prt_z \varphi_1)^2 : - \frac{1}{4} : (\prt_z
\varphi_2)^2: + i \al_0 \prt^2_z \varphi_1 \nonumber \\
\frac{i(22+5c)}{4} W_3 &=& : (\prt_z \varphi_2)^3:-: (\prt_z \varphi_1)^2
\prt_z \varphi_2 : + i \al_0 : \prt^2_z \varphi_1 \prt_z \varphi_2: \nonumber
\\
&& + 3i \al_0 : \prt_z \varphi_1 \cdot \prt^2_z \varphi_2 : +2 \al_0^2
\prt^3_z \var_2
\ena
where
\be
<\varphi_a(z) \cdot \varphi_b(w)>=-2 \delta_{ab} \ \ln (z-w)
\ee
using the results of \cite{1}.

Denoting $\vec{\alpha}$ and $\vec{\beta}$ the 2-dim. simple roots of $Sl(3)$,
one can reexpress the two components $(\varphi_1, \varphi_2)$ of $\vec{\phi}$
in the
$(\vec{\al}, \vec{\beta})$ basis:
\bea
\var_1 &=& \frac{1}{3} (2\varphi_\al + \var_\beta) \nonumber \\
\var_2 &=& \frac{1}{3} (-\var_\al - \var_\beta)
\ena
with: $\var_\al = \vec{\phi} \cdot \vec{\al}$ and $\var_\beta = \vec{\phi}
\cdot \vec{\beta}$.

As an aside remark, we recognize in the $i\al_0 \prt^2_z \var_1$ "screening"
term of $W_2$ the usual quantity $i\al_0 \vec{\rho} \cdot \prt^2_z
\vec{\phi}$ where $\vec{\rho}$ is the half sum of the positive roots, that is
here $\ \ \shalf [ \vec{\al} + \vec{\beta} + (\vec{\al} + \vec{\beta})]
= \vec{\al} + \vec{\beta}$.

Then, one sees immediately that the condition:
\be
\var_\al = \var_\beta
\ee
corresponding to the folding $Sl(3) \rightarrow SO(3)$ implies
\be
W_3 (\var_\al = \var_\beta)=0
\ee
as could be expected.

Actually, one knows that, in the $Sl(n)$ case, the fields $W_{h+1}(z)$
can be fully
determined by considering the formal differential operator \cite{2}:
\be
R_n = : \prod^n_{j=1} \left( i \al_0 \frac{d}{dz} +
\frac{\vec{h}_j}{\sqrt{2}} \prt_z \vec{\phi} \right):
\ee
where $\vec{\phi} = \left( \var_1, ... \var_{n-1} \right)$ and the vectors
$\vec{h}_i$ with $i=1,2,...,n$ are normalized such that
\be
\vec{h}_i \vec{h}_j = \delta_{ij} - \frac{1}{n} \ \ \ \ \ \ \ \sum^n_{i=1}
\vec{h}_i = \vec{0};
\ee
and setting:
\be
R_n = \sum^n_{k=0} W_k(z) \cdot \left( i \al_0 \frac{d}{dz}
\right)^{n-k}
\ee

the transformation from the fields $\prt_z \var_i$ to the $W_k(z)$ ones being
called the Miura transformation.

Let us define, as usual, the simple roots of $Sl(n)$ as:
\bea
\vec{\al_1} &=& \vec{h_1} - \vec{h_2} \nonumber \\
\vdots & \nonumber \\
\vec{\al}_{n-1} &=& \vec{h}_{n-1} - \vec{h}_n
\ena
and let us express the $\vec{h}_i$'s in terms of the $\vec{\al_i}$'s:
\bea
\vec{h}_1 &=& \frac{1}{n} \left[ (n-1) \vec{\al}_1 + (n-2) \vec{\al_2} + ... +
2 \vec{\al}_{n-2} + \vec{\al}_{n-1} \right] \nonumber \\
\vec{h}_2 &=& \frac{1}{n} \left[ -\vec{\al}_1 + (n-2) \vec{\al_2} + ... +
\vec{\al}_{n-1} \right] \nonumber \\
\vdots && \nonumber \\
\vec{h}_k &=& \frac{1}{n} \left[ -\vec{\al}_1 - 2\vec{\al_2} - ... -
(k-1) \vec{\al}_{k-1} + (n-k)\vec{\al}_k + ... + \vec{\al}_{n-1}
\right] \nonumber \\
\vdots && \nonumber \\
\vec{h}_{n-1} &=& \frac{1}{n} \left[ -\vec{\al}_1 ...
- (n-2)  \vec{\al}_{n-2} + \vec{\al}_{n-1} \right] \nonumber \\
\vec{h}_n &=& \frac{-1}{n} \left[ \vec{\al}_1 + ... +
(n-1) \vec{\al}_{n-1} \right]
\ena

Then, one remarks, by identifying once more:
\be
\vec{\al_i} \cdot \prt_z \vec{\phi} = \vec{\al}_{n-i} \cdot \prt_z \vec{\phi}
\ \ \ \ \ \ \ \ i=1,...,n-1.
\ee
that
\be
\vec{h}_j \cdot \prt_z \vec{\phi} = - \vec{h}_{n+1-j} \cdot \prt_z \vec{\phi}
\ \ \ \ \ \ \ j=1,2,...,n
\ee
and when $n = 2p+1$
\be
\vec{h}_{p+1} \cdot \prt_z \vec{\phi} =0.
\ee

Therefore the $R$ differential operator becomes in the $C_p$ case $(n=2p)$:
\be
\left( i \al_0 \frac{d}{dz} + \prt_z \tilde{\phi}_1 \right)
\, ...\, \left(
i \al_0 \frac{d}{dz} + \prt_z \tilde{\phi}_p \right) \left( i \al_0
\frac{d}{dz} - \prt_z \tilde{\phi}_p \right)
\, ...\, \left( i \al_0 \frac{d}{dz} - \prt_z \tilde{\phi}_1 \right)
\ee
and in the $B_p$ case $(n=2p+1)$:
\be
\left( i \al_0 \frac{d}{dz} + \prt_z \tilde{\phi}_1 \right)\, ...\,
\left(
i \al_0 \frac{d}{dz} + \prt_z \tilde{\phi}_p \right) \left(i\al_0
\frac{d}{dz} \right) \left( i \al_0
\frac{d}{dz} - \prt_z \tilde{\phi}_p \right)
\, ...\,
\left( i \al_0 \frac{d}{dz} - \prt_z \tilde{\phi}_1 \right)
\ee
where by $\prt_z \tilde{\phi}_i$ we mean the quantity $\frac{1}{\sqrt{2}}
\vec{h}_i \cdot \prt_z \vec{\phi}$ in which are identified $\var_{\al_i}$ and
$\var_{\al_{n-i}}$.

These two last expressions are exactly the ones given in \cite{3} or
\cite{4} for the Miura transformation in the $C_n$ and $B_n$ case !

\sect{The Supersymmetric case \label{sec:9}}

\subsection{Folding of superalgebras \label{subsec:1}}

\indent

Let $\cg$ be a simple Lie superalgebra. Then $Out (\cg)$ is isomorphic
to $\Z_2$
for $A(m,n)=Sl(m+1|n+1)$ with $m \neq n \neq 0, A(1,1), A(0,2n+1)$ and
$D(m,n)=OSp(2m|2n)$ with $m \neq 1,$ to $\Z_2 \times \Z_2$
for $A(m,n)$ with
$m \neq 0,1$ and to $\Z_4$ for $A(0,2n)$. It reduces to the identity for
$B(m,n)=OSp(2m+1|2n), C(n+1)=OSp(2|2n), F(4)$ and $G(3)$.

In analogy with the algebraic case, outer automorphisms of a superalgebra are
related to the symmetries of the corresponding Dynkin diagrams (DD) (notice
that
there is in general many inequivalent DD's for a superalgebra $\cg$). When $Out
(\cg)$ does not reduce to the identity, that means that it exists at least one
DD of $\cg$, the symmetry group of which is isomorphic to it. Using the
symmetries of the DD's of simple Lie superalgebras \cite{11}, one finds the
schemes summarized in Table 1.

We note that all the foldings $Sl \rightarrow OSp$ are of the type $Sl(M|2n)
\rightarrow OSp (M|2n)$. Considering the $(M+N) \times (M+N)$
matrix representation of $Sl(M|N)$ (with $N=2n$ even), the folding $Sl(M|N)
\rightarrow OSp(M|N)$ is associated to symmetries of the constituting blocks of
the matrices. More precisely, if an element of $Sl(M|N)$ is denoted by
\be
E= m^{ij} E_{ij} \ \ \ \mbox{with} \ m^{ij} \ \ \mbox{real numbers}
\ee
submitted to the supertraceless condition
\be
\sum^M_{i=1} m^{ii} - \sum^N_{j=1} \ m^{M+j, M+j} =0,
\ee
a representation of $OSp(M|N)$ is obtained with the matrices $E$ with
coefficients $m^{ij}$ satisfying the following relations:

$i)$ if $M$ is odd ($M=2m+1$ and $N=2n$), one has
\be
m^{ij} = (-1)^{i+j+1} m^{M+1-j,M+1-i} \ \ \ \ \mbox{with } 1 \le i,j \le M
\ee
\be
m^{M+i,M+j} = (-1)^{i+j+1} m^{M+N+1-j,M+N+1-i} \ \ \ \ \mbox{with }
1 \le i,j \le N
\ee
\be
m^{i,M+j} = (-1)^{m+n+i+j} m^{M+N+1-j,M+1-i} \ \ \ \ \mbox{with }
1 \le i \le M \mbox{ and } 1 \le j \le  N
\ee

\newpage

Page of pictures

\newpage
$ii)$ if $M$ is even ($M=2m$ and $N=2n$), one has
\be
m^{i,j} = (-1)^{i+j+\epsilon (i)+\epsilon (M+1-j)} m^{M+1-j,M+1-i}
\ \ \ \ \ \mbox{with } 1 \le i,j \le M
\ee
\be
m^{M+i,M+j} = (-1)^{i+j+1} m^{M+N+1-j,M+N+1-i} \ \ \ \ \ \mbox{with }
1 \le i,j \le N
\ee
\be
m^{i,M+j} = (-1)^{m+n+i+j+\epsilon (i)} m^{M+N+1-j,M+1-i}
\ \ \ \ \ \mbox{with } 1 \le i \le M \mbox{ and } 1 \le j \le N
\ee
where $\epsilon (i) = \left[\frac{i}{m+1}\right]$.

For example, an element of $Sl(3|2)$ being given, a general element of
$OSp(3|2)$ will write
\be
\left(
\begin{array}{ccc|cc}
x & a & 0 & m  & n \\
b & 0 & a & p & q \\
0 & b & -x & r & s \\
\hline
-s & q & -n & x+y & c \\
r & -p & m & d & -x-y
\end{array}
\right)
\label{eq:luc}
\ee

\subsection{Folding of a $W$ superalgebra \label{subsec:2}}

\indent

In order to perform the folding of super $W$ algebras induced by the folding of
superalgebras (SA), the method one has to follow is completely the same as for
the bosonic case. In particular, the determination of the $OSp(1|2)$
subsuperalgebras  which will  survive the folding is given by looking at the
decomposition with each $OSp(1|2)$ of the fundamental representation
$\underline{M+N}$ of $\cg = Sl(M|N)$ and $\cg^F = OSp(M|N)$
respectively. Each time one finds the same decomposition for \cg\ and $\cg^F$,
the corresponding $OSp(1|2)$ is a superprincipal embedding either of \cg \ or
of
a subsuperalgebra (SSA) $\tilde{\cg}$ of $\cg$. For example, in the case of
$Sl(3|2)$, one has:

\vs{2}

\begin{tabular}{ccc}
$OSp(1|2)$ & $Sl(3|2)$ regular SSA where
& $OSp(3|2)$ regular \\
decomposition \hs{14}& $OSp(1|2)$ is superprincipal \hs{14} & SSA \\
&& \\
$\r{1}$ & $Sl(3|2)$ & $OSp(3|2)$ \\
$\r{1/2}+ \rpi{0} + \r{0}$ & $Sl(2|1)$ & no counterpart \\
$\rpi{1/2} + 2\r{0}$ & $Sl(1|2)$ & $\left\{ \begin{array}{c} OSp(1|2) \\
OSp(2|2) \end{array} \right.$ \\
\end{tabular}

\vs{3}

We consider in some detail the two different $OSp(3|2)$ models arising
from the folding $Sl(3|2)\rightarrow OSp(3|2)$ in appendix A.
We recall that an irreducible representation of $OSp(1|2)$ is the direct sum
of two $Sl(2)$ representations ${\cal{D}}_j \oplus {\cal{D}}_{j-1/2}$ ($j$
integer or half-integer) with an exception for
the trivial one-dimensional representation ${\cal{D}}_0=\r{0}$.
Moreover, in the decomposition of the superalgebra $Sl(m|n)$ under $OSp(1|2)$,
we denote the $OSp(1|2)$ representations by $\r{j}$ if the
representation ${\cal{D}}_j$ comes from the decomposition of the
fundamental of $Sl(m)$ and $\rpi{j}$ if ${\cal{D}}_j$ comes from the
decomposition of the fundamental of $Sl(n)$.

One has to remark that, unlike the bosonic case, the folding cannot be
performed for Abelian theories in all cases since among the unitary SA's to be
folded, only the $Sl(2n \pm 1|2n)$ superalgebras lead to Abelian Toda models.

Therefore, only the $OSp(2n \pm 1|2n)$ Abelian models can be obtained  by
simple folding. For the cases $OSp(2n|2n)$ and $OSp(2n+2|2n)$ (Abelian theory)
one has
to take a more complicated way as explained in subsections
\ref{subsec:4} and \ref{subsec:5}.

\subsection{An example : $OSp(1|2)$-$W$ superalgebra from $Sl(2|1)$ one
\label{subsec:3}}

\indent

The simplest example comes from the folding $Sl(2|1) \rightarrow OSp(1|2)$. All
the needed material is furnished in section 6 of reference \cite{12}. Indeed
the constrained current matrix reads in the $Sl(2|1)$ case :
\be
J^g_{Sl(2|1)} =
\left(
\begin{array}{cc|c}
\phi_1 & 0 & \mu_2 \\
Y & \phi_2 & U_2 \\
\hline
U_1 & \mu_1 & \phi_1 + \phi_2
\end{array}
\right)
\ee
with $\phi_1, \phi_2, Y$ are fermionic superfields, $U_1, U_2$ bosonic
superfields and finally $\mu_1$ and $\mu_2$ constants involved in the
constraints. The $OSp(1|2)$ current matrix is:
\be
J^g_{OSp(1|2)} =
\left(
\begin{array}{cc|c}
\phi & 0 & \mu \\
Y & - \phi & -U \\
\hline
U & \mu & 0
\end{array}
\right)
\ee

Then, imposing:
\bea
\phi_1 &=& -\phi_2 = \phi \nonumber \\
U_1 &=& -U_2 = U \nonumber \\
\mu_1 &=& \mu_2 = \mu
\ena
in the $\wt{W}$ superfield expressions\footnote{We denote by $\wt{W}$ the
superfield generators to distinguish them from the fields $W$.}
(cf. eq.(6.21) of \cite{12})
\bea
\wt{W}_{3/2} &=& \frac{1}{2 \mu_1 \mu_2} \left(2 \mu_1 \mu_2 Y - 2 \mu_2 \phi_2
U_1
-2 \mu_1 \phi_1 U_2 - \phi_1 D\phi_2 -\phi_2 D\phi_1 \right. \nonumber \\
&& \left. + \mu_2 DU_1 - \mu_1 DU_2 + D^2(\phi_1-\phi_2) \right) \nonumber \\
\wt{W}_1 &=& \frac{1}{2} \left(\mu_2 U_1 + \mu_1 U_2 + \phi_1 \phi_2 + D\phi_1
+D\phi_2 \right)
\ena
one obtains $\wt{W}_1=0$ (as expected) and:
\be
\wt{W}_{3/2}^F = \frac{1}{\mu^2} \left( \mu^2 Y +2 \mu \phi U
+\phi D\phi +\mu DU +D^2\phi \right)
\ee
this last expression being exactly the one obtained in \cite{12} from the
$OSp(1|2)$ superalgebra.

\subsection{The case of $OSp(2n|2n)$ \label{subsec:4}}

\indent

Although the $OSp(2n|2n)$ is obtained by folding of the $Sl(2n|2n)$
superalgebra,
one cannot use it for the corresponding ${W}$ algebras, since $Sl(2n|2n)$ does
not admit an Abelian Toda theory ($Sl(2n|2n)$ has an odd simple root system but
one cannot find a superprincipal $OSp(1|2)$ in $Sl(2n|2n)$).

The method we will use will be to consider first the folding $Sl(2n+ 1|2n)
\Rightarrow OSp(2n+1|2n)$ and then to pick in $OSp(2n+1|2n)$ the regular
$OSp(2n|2n)$ SSA (in the same way one finds $D_n$ from $A_{2n}$ via $B_n$ in
the bosonic case).

\indent

Indeed, the $OSp(2n|2n)$ fundamental representation decomposes under the
superprincipal $OSp(1|2)$ as

\be
\underline{4n} = \rpi{n-1/2} + \r{0}.
\ee

In order to find the Abelian model associated to $OSp(2n|2n)$ starting from
$Sl(2n+1|2n)$, one has to consider the $OSp(1|2)$ embedding in $Sl(2n+1|2n)$
such that the decomposition of the fundamental representation of $Sl(2n+1|2n)$
takes the form
\be
\underline{4n+1} = \rpi{n-1/2} + 2\r{0}.
\ee

This is the case if and only if the regular SSA of $Sl(2n+1|2n)$ where
$OSp(1|2)$ is superprincipal is $Sl(2n-1|2n)$.

The superdefining vector of the embedding $OSp(1|2)_{ppal} \subset Sl(2n-1|2n)
\subset Sl(2n+1|2n)$ is
given by
\be
\left( n-1,...,1,0,0,0,-1,...,-n+1;n-1/2,n-3/2,...,-n+1/2,-n+3/2 \right)
\ee
which leads to the following grading matrix
\be
\hs{-10}
\left(
\begin{array}{ccccccccccc|ccccc}
0 &1 &2 &\cdots & \sma{n-1} & \sma{n-1} & \sma{n-1} &n &\cdots &\cdots
& \sma{2n-2} &-\frac{1}{2} &\cdots && \cdots & \sma{2n-\frac{3}{2}} \\
  &\ddots &\ddots &\ddots & \vdots & \vdots & \vdots
&\ddots &\ddots & &\vdots &\vdots &&&&
\vdots \\
  & &0 &1 &2 &2 &2 & &\ddots &\ddots &\vdots &\vdots &&&& \vdots \\
  & & &0 &1 &1 &1 &2 & &\ddots &n & \sma{-n+\frac{3}{2}} &\cdots && \cdots &
\sma{n+\frac{1}{2}} \\
  & & & &0 &0 &0 &1 &2 &\cdots & \sma{n-1} & \sma{-n+\frac{1}{2}} &\cdots &&
\cdots &
\sma{n-\frac{1}{2}} \\
  & & & &0 &0 &0 &1 &2 &\cdots & \sma{n-1} & \sma{-n+\frac{1}{2}} &\cdots &&
\cdots &
\sma{n-\frac{1}{2}} \\
  & & & &0 &0 &0 &1 &2 &\cdots & \sma{n-1} & \sma{-n+\frac{1}{2}} &\cdots &&
\cdots &
 \sma{n-\frac{1}{2}} \\
  & & & & & & &0 &1 &\cdots & \sma{n-2} & \sma{-n-\frac{1}{2}} &\cdots &&
\cdots &
\sma{n-\frac{3}{2}}\\
  & & & & & & & &\ddots &\ddots &\vdots &\vdots &&&& \vdots \\
  & & & & & & & & &0 &1 &\vdots &&&& \vdots \\
  & & & & & & & & & &0 & \sma{-2n+\frac{3}{2}} &\cdots && \cdots &
+\frac{1}{2}\\
\hline
\frac{1}{2}  & \cdots & & \cdots & \sma{n-\frac{3}{2}} & \sma{n-\frac{1}{2}} &
\sma{n-\frac{1}{2}} &
\sma{n-\frac{1}{2}} & \sma{n+\frac{1}{2}}& \cdots & \sma{2n-\frac{3}{2}} & 0 &1
&2 &\cdots &
\sma{2n-1} \\
\vdots  & & & & \vdots & \vdots & \vdots & \vdots & \vdots &
& \vdots & &\ddots &\ddots &\ddots
&\vdots \\
  & & & & & & & & & & & & &0 &1 &2 \\
\vdots  & & & & \vdots & \vdots & \vdots & \vdots & \vdots &
& \vdots & & & &0 &1 \\
\sma{\frac{3}{2}-2n} & \cdots & & \cdots & \sma{-n-\frac{1}{2}} &
\sma{\frac{1}{2}-n} &
\sma{\frac{1}{2}-n} & \sma{\frac{1}{2}-n} & \sma{\frac{3}{2}-n} &
\cdots & -\frac{1}{2} & &
& & & 0 \\
\end{array}
\right)
\ee

\indent

The superspin content is

\be
\left\{
\begin{array}{l}
\wt{W}_{2n-1}, \wt{W}_{2n-2},...,\wt{W}_1 \nonumber \\
\wt{W}_{2n-1/2}, \wt{W}_{2n-3/2},...,\wt{W}_{3/2}, 4\wt{W}_{1/2} \nonumber \\
4\wt{W}_n
\end{array}
\right.
\ee

Now, performing the folding $Sl(2n+1|2n) \rightarrow OSp(2n+1|2n)$, one finds
the superspin content corresponding to the $OSp(2n+1|2n)$ Toda theory, i.e.
\be
\left\{
\begin{array}{l}
\wt{W}_{2n-2}, \wt{W}_{2n-4},...,\wt{W}_2 \nonumber \\
\wt{W}_{2n-1/2}, \wt{W}_{2n-5/2},...,\wt{W}_{3/2}, 3\wt{W}_{1/2} \nonumber \\
3\wt{W}_n
\end{array}
\right.
\ee

To obtain the Toda model based on $OSp(2n|2n)$ from $OSp(2n+1|2n)$, one
suppresses the row $N^{th}$ and the $N^{th}$ column in the matrix
representation, with $N=2n+1$, leading to the superspin content of the
$OSp(2n|2n)$ Abelian super-Toda theory:
\be
\left\{
\begin{array}{l}
\wt{W}_{2n-2}, \wt{W}_{2n-4},...,\wt{W}_2 \nonumber \\
\wt{W}_{2n-1/2}, \wt{W}_{2n-5/2},...,\wt{W}_{3/2} \nonumber \\
\wt{W}_n
\end{array}
\right.
\ee

\subsection{The $OSp(2n+2|2n)$ Abelian Toda case \label{subsec:5}}

\indent

Finally, let us consider the case of $OSp(2n+2|2n)$. This superalgebra  can be
obtained by folding $Sl(2n+2|2n)$. Although this last superalgebra does not
admit a superprincipal $OSp(1|2)$, it is still possible to build the Abelian
theory for $OSp(2n+2|2n)$ by folding a suitable model for $Sl(2n+2|2n)$.

Indeed, the Abelian theory for $OSp(2n+2|2n)$ corresponds to the following
decomposition of $OSp(2n+2|2n)$ under its superprincipal $OSp(1|2)$:
\be
\underline{4n+2} = \r{n}\oplus \r{0}
\ee
Now, looking at the $Sl(2n+2|2n)$ models, such a decomposition for the
fundamental representation of $Sl(2n+2|2n)$ is obtained when
one considers an $OSp(1|2)$
superprincipal in the regular SSA $Sl(2n+1|2n)$. The corresponding superspin
content is given by
\be
\wt{W}_{2n+1/2}, \wt{W}_{2n},...,\wt{W}_{1/2}, 2\wt{W}_{n+1/2}
\ee

Then, performing the folding, the fields $\wt{W}_{2n+1/2},
\wt{W}_{2n-3/2},...,\wt{W}_{2n-1},...,\wt{W}_1,\wt{W}_{n+1/2}$ are killed,
leading to the Abelian
Toda $OSp(2n+2|2n)$ theory with superspin content
\be
\left\{
\begin{array}{l}
\wt{W}_{2n}, \wt{W}_{2n-2},...,\wt{W}_2 \\
\wt{W}_{2n-1/2},\wt{W}_{2n-5/2},...,\wt{W}_{3/2} \\
\wt{W}_{n+1/2}
\end{array}
\right.
\ee

Taking for illustration the $Sl(4|2)$ case, one obtains for the model with
$\tilde{\cg} =Sl(3/2)$:
\be
J^g=
\left(
\begin{array}{cccc|cc}
\wt{W}_{1/2} & \wt{W}_{3/2} & 4\wt{W}''_{3/2} - \wt{W}_{3/2} & \wt{W}_{5/2}
& 4\wt{W}_1 & \wt{W}_2 \\
0 & 0 & \wt{W}_{1/2} & \wt{W}'_{3/2} & 1 & \wt{W}_1 \\
0 & \wt{W}_{1/2}  & 0 & 4\wt{W}''_{3/2}-\wt{W}'_{3/2} & 1 & \wt{W}_1 \\
0 & 0 & 0 & \wt{W}_{1/2} & 0 & 1 \\
\hline
1 & \wt{W}_1 & \wt{W}_1 & \wt{W}_2 & \wt{W}_{1/2} & \wt{W}''_{3/2} \\
0 & 1 & 1 & 4\wt{W}_1 & 0 & \wt{W}_{1/2}
\end{array}
\right)
\ee

The superspin content is given by $\wt{W}_{5/2}, \wt{W}_2, \wt{W}_{3/2},
\wt{W}'_{3/2}, \wt{W}''_{3/2},
\wt{W}_1, \wt{W}_{1/2}$. The folding leads to the following $J^g$ current for
the
$OSp(4|2)$ Abelian model:
\be
J^g=
\left(
\begin{array}{cccc|cc}
0 & \wt{W}_{3/2} & \wt{W}'_{3/2} & 0 & 0 & \wt{W}_2 \\
0 & 0 & 0 & \wt{W}'_{3/2} & 1 & 0 \\
0 & 0 & 0 & \wt{W}_{3/2} & 1 & 0 \\
0 & 0 & 0 & 0 & 0 & 1 \\
\hline
1 & 0 & 0 & \wt{W}_2 & 0 & \sm{1}{4}(\wt{W}_{3/2} + \wt{W}'_{3/2} ) \\
0 & 1 & 1 & 0 & 0 & 0
\end{array}
\right)
\ee
where the fields $\wt{W}_{5/2}, \wt{W}_1, \wt{W}_{1/2}$ have been
killed and $\wt{W}''_{3/2} =
\sm{1}{4}(\wt{W}_{3/2} + \wt{W}'_{3/2})$, leaving the superspin content
$\wt{W}_2, \wt{W}_{3/2},
\wt{W}'_{3/2}$ as expected.

\vs{12}

\indent

{\Large{\bf {Acknowledgements}}}

\indent

We are indebted to K. Hornfeck for crucial discussions and informations on the
computation of fusion rules.

\vs{14}
\appendix

\sect{Appendix: $OSp(3|2)$-$W$ algebras from $Sl(3|2)$ ones}

\indent

The folding $Sl(3|2) \rightarrow OSp(3|2)$ provides two different models, since
$Sl(3|2)$ contains (see section \ref{subsec:2})
two different $OSp(1|2)$, each
being superprincipal in a regular subsuperalgebra $\tilde{\cg}$.

\indent

(i) Case $\tilde{\cg} = Sl(3|2)$

The fundamental representation of $Sl(3|2)$ decomposes as
$\underline{5} = \r{1}$.

The superdefining vector of the embedding is therefore $(1,0,-1;\shalf,
-\shalf)$, and
the grading is given by the following antisymmetric matrix:
\be
g=
\left(
\begin{array}{ccc|cc}
0 &1&2&1/2&3/2 \\
-1&0&1&-1/2&1/2\\
-2&-1&0&-3/2&-1/2\\
\hline
-1/2&1/2&3/2&0&1\\
-3/2&-1/2&1/2&-1&0
\end{array}
\right)
\ee

The constraints on the currents being given by
\be
J_{-1/2} = \sum^4_{i=1} \ \ \mu_i E_{-\al_i}
\ee
$E_{-\al_i}\ i=1,..,4$ are the generators associated to
the negative simple roots $-\al_i$, one has to choose the coefficients $\mu_i$
compatible with the folding. Comparing with the matrix (\ref{eq:luc}), we
choose
\be
\left\{
\begin{array}{l}
\mu_{2k}=-1\\
\mu_{2k-1}=1
\end{array}
\right.
\ \ \mbox{ with } \ \ 1 \leq k \leq 2
\ee

Then the superprincipal $OSp(1|2)$ is generated by
\bea
M_{-1/2} &=& E_{41} - E_{24} + E_{52} - E_{35}  \nonumber \\
M_{+1/2} &=& 2E_{14} + E_{42} + E_{25} +2 E_{53} \nonumber \\
M_0 &=& E_{11} - E_{33} + \shalf E_{44} - \shalf E_{55} \nonumber \\
M_\pm &=& \shalf \left\{ M_{\pm 1/2} , M_{\pm1/2} \right\}
\ena

One determines the $Sl(3|2)$ highest weights $X$ under the superprincipal
$OSp(1|2)$ by imposing
\bea
\{ M_{+1/2},X_F \} &=& 0 \nonumber \\
\mbox{and } \ \left[ M_{+1/2} , X_B \right] &=&0
\ena
where $X_F$ are fermionic and $X_B$ bosonic.

One finally finds the constrained current $J^g$:
\be
J^g =
\left(
\begin{array}{ccc|cc}
0 &2\wt{W}_{3/2}&\wt{W}_{5/2}&2\wt{W}_1&\wt{W}_2 \\
0&0&2\wt{W}_{3/2} &-1&\wt{W}_1 \\
0&0&0&0&-1\\
\hline
1&-\wt{W}_1&-\wt{W}_2&0&\wt{W}_{3/2} \\
0&1&-2\wt{W}_1&0&0
\end{array}
\right)
\ee
which leads to $\wt{W}$ fields with superspins $5/2, 3/2, 2,1$. Applying the
symmetry (\ref{eq:luc}), one obtains the corresponding current $J^g$ for
$OSp(3|2)$, i.e.
\be
J^g =
\left(
\begin{array}{ccc|cc}
0 &2\wt{W}_{3/2}&0&0&\wt{W}_2 \\
0&0&2\wt{W}_{3/2} &-1& 0 \\
0&0&0&0&-1\\
\hline
1&0&-\wt{W}_2&0&\wt{W}_{3/2} \\
0&1&0&0&0
\end{array}
\right)
\ee

One finds as expected that the spin $5/2$ and the spin 1 fields disappear.

\indent

(ii) Case $\tilde{\cg} = Sl(1|2)$

The fundamental representation of $Sl(3|2)$ decomposes as $\underline{5}
= \rpi{1/2} + 2 \r{0}$
and the $OSp(1|2)$, superprincipal in $\tilde{\cg}$,  is generated by:
\bea
M_{-1/2} &=& E_{14} + E_{24} + E_{34} + E_{51} - E_{52} + E_{53} \nonumber \\
M_{+1/2} &=& -E_{15} - E_{25} - E_{35} + E_{41} - E_{42} + E_{43} \nonumber \\
M_0 &=& \shalf (E_{44} - E_{55}) \nonumber \\
M_\pm &=& \shalf \left\{ M_{\pm 1/2} , M_{\pm 1/2} \right\}
\ena
The superdefining vector of the embedding is given by $(0,0,0;\shalf, -
\shalf)$ and the grading matrix is
\be
g=
\left(
\begin{array}{ccc|cc}
0 & 0& 0& -1/2& 1/2 \\
0 & 0& 0& -1/2& 1/2 \\
0 & 0& 0& -1/2& 1/2 \\
\hline
1/2 & 1/2 & 1/2 & 0 & 1 \\
-1/2 & -1/2 & -1/2 & -1 & 0
\end{array}
\right)
\ee

The $J^g$ matrix for $Sl(3|2)$ is then given by:
\be
J^g=
\left(
\begin{array}{ccc|cc}
\wt{W}'_{1/2} + \wt{W}''_{1/2} & \wt{W}_{1/2} & \wt{W}'_{1/2} & 1
& \wt{W}^{(1)}_1 \\
\wt{W}''_{1/2} & \wt{W}_{1/2}+2\wt{W}'_{1/2}  & \wt{W}'''_{1/2} & 1
& \wt{W}^{(2)}_1 \\
 & -\wt{W}'''_{1/2} & & & \\
\wt{W}_{1/2} + \wt{W}'_{1/2} & -\wt{W}_{1/2}-\wt{W}''_{1/2}
& \wt{W}_{1/2}+\wt{W}'_{1/2} & 1 & \wt{W}^{(3)}_1 \\
 &+ \wt{W}''_{1/2}   & + \wt{W}''_{1/2} + \wt{W}'''_{1/2} & & \\
\hline
\wt{W}^{(4)}_1 & \wt{W}^{(5)}_1 & \al & \wt{W}_{1/2} + 2\wt{W}'_{1/2}
& \wt{W}_{3/2} \\
& & & + \wt{W}''_{1/2} & \\
1 & -1 & 1 & 0 & \wt{W}_{1/2} + 2\wt{W}'_{1/2} \\
 & & & & + \wt{W}''_{1/2}
\end{array}
\right)
\ee
$(\al = \wt{W}^{(1)}_1 - \wt{W}^{(2)}_1 + \wt{W}^{(3)}_1 -
\wt{W}^{(4)}_1 + \wt{W}^{(5)}_1 )$

\indent

We obtain as expected one superspin 3/2, five superspin 1 and four superspin
1/2 fields.

\indent

Now applying again the symmetry equations (\ref{eq:luc}), one has
the $J^g$ current matrix for $OSp(3|2)$:
\be
J^g=
\left(
\begin{array}{ccc|cc}
-\wt{W}_{1/2} & \wt{W}_{1/2} & 0 & 1 & -\wt{W}'_1 \\
-\wt{W}_{1/2} & 0 & \wt{W}_{1/2} & 1 & \wt{W}_1 \\
0 & -\wt{W}_{1/2} & \wt{W}_{1/2} & 1 &\wt{W}'_1 \\
\hline
-\wt{W}'_1 & \wt{W}_1 & \wt{W}'_1 & 0 & \wt{W}_{3/2} \\
1 & -1 & 1 & 0 & 0
\end{array}
\right)
\ee

One finds the superspin 3/2, two superspin 1 and only one superspin 1/2 field,
in
accordance with Table 11 of reference \cite{7}.


\newpage

\begin{thebibliography}{99}
\bibitem{1b}      D.~Olive and N.~Turok, \NP{B215} (1983) 47.
\bibitem{2b}      P.~Goddard, W.~Nahm, D.I.~Olive, A.~Schrimmer,
                  \CMP{107} (1986) 179.
\bibitem{3b}      L.~Frappat, A.~Sciarrino and P.~Sorba, \NP{B305} (1988) 164.
\bibitem{4b}      L.~Frappat, A.~Sciarrino and P.~Sorba, \JMP{30} (1989) 2984.
\bibitem{5b}      For a recent review, see P.~Bouwknegt and K.~Schoutens,
                  {\em "$W$ symmetry in conformal field theory"},
                  preprint CERN-TH-6583/92, ITP-SB-92-23, to be published
                  in Physics Reports.
\bibitem{6b}      L.~Feher, L.~O'Raifeartaigh, P.~Ruelle, I.~Tsutsui,
                  and A.~Wipf,\\
                  {\em "On the general structure of hamiltonian reductions
                  of the WZNW theory"},\\
                  Physics Report {\bf 222} (dec. 1992) $n^o$ 1,
                  and references therein.
\bibitem{7}       L.~Frappat, E.~Ragoucy and P.~Sorba, {\em "$W$ algebras and
                  superalgebras from constrained WZW models:
                  a group theoretical classification"},
                  preprint ENSLAPP-AL-391/92.
\bibitem{H1}      K. Hornfeck, \PL{275B} (1992) 355.
\bibitem{KW}      H.G. Kausch and G.M.T. Watts, \NP{B354} (1991) 740.
\bibitem{H2}      K. Hornfeck, private communication.
\bibitem{1}       V.A. Fateev and AB Zamolodchikov, \NP{B280} [FS18] (1987)
644.
\bibitem{2}       V.A. Fateev and S,.L. Lukyanov, Int. Journ. of Mod. Phys.
                  {\bf A}Vol. 3 (1988) 507.
\bibitem{3}       V.A. Fateev and S,.L. Lukyanov, Landau preprint,
                  Moscow (1988),\\
                  {\em "Additional symmetries and exactly soluble models
                  in 2 dim.
                  conformal field theory I: Quantization and Hamiltonian
                  structures"}.
\bibitem{4}       A. Bilal and J.-L. Gervais, \NP{B314} (1989) 646.
\bibitem{11}      L.~Frappat, A.~Sciarrino and P.~Sorba, \CMP{121} (1989) 457.
\bibitem{12}      F.~Delduc, E.~Ragoucy and P.~Sorba, \CMP{146} (1992) 403.
\end{thebibliography}
\end{document}